\documentclass[aps,prl,reprint,superscriptaddress,amsmath,floatfix]{revtex4-1}
\usepackage{graphicx}
\usepackage{siunitx}
\usepackage{bm}
\usepackage{color}
\usepackage[T1]{fontenc}
\usepackage[normalem]{ulem}      

\begin{document}

\title{Coupling between Inclusions and Membranes at the Nanoscale}
\author{Florent Bories}
\affiliation{Laboratoire ``Mati\`{e}re et Syst\`{e}mes Complexes'' (MSC), UMR 7057 CNRS, Universit\'{e}
Paris 7 Diderot, 75205 Paris Cedex 13, France}
\author{Doru Constantin}
\email{doru.constantin@u-psud.fr} \homepage{www.equipes.lps.u-psud.fr/constantin/}
\affiliation{Laboratoire de Physique des Solides, CNRS, Univ. Paris-Sud,
Universit\'{e} Paris-Saclay, 91405 Orsay Cedex, France}
\author{Paolo Galatola}
\email{paolo.galatola@univ-paris-diderot.fr.}
\author{Jean-Baptiste Fournier}
\email{jean-baptiste.fournier@univ-paris-diderot.fr.}
\affiliation{Laboratoire ``Mati\`{e}re et Syst\`{e}mes Complexes'' (MSC), UMR 7057 CNRS, Universit\'{e} Paris 7 Diderot, 75205 Paris Cedex 13, France}

\date{\today}
			
\begin{abstract}
The activity of cell membrane inclusions (such as ion channels) is
influenced by the host lipid membrane, to which they are elastically
coupled. This coupling concerns the hydrophobic thickness of the bilayer
(imposed by the length of the channel, as per the hydrophobic
matching principle) but also its slope at the boundary of the
inclusion. However, this parameter has never been measured so far. We
combine small-angle x-ray scattering data and a complete elastic model
to measure the slope for the model gramicidin channel and show that it
is surprisingly steep, in two membrane systems with very different
elastic properties. This conclusion is confirmed and generalized by the
comparison with recent results in the simulation literature and with
conductivity measurements.
\end{abstract}
				
\maketitle

The coupling with the lipid membrane plays an important role in the activity of membrane-bound proteins~\cite{Gil:1998,Lee:2005}, and a sustained research effort aims to describe this interaction, either in the framework of continuum theories, or at the microscopic level.

A very reliable conclusion has been that, if the hydrophobic length of the trans-membrane domain of a protein is different from the hydrophobic thickness of the surrounding membrane, the latter is ``pinched'' or ``stretched'' to adapt to the (much more rigid) protein~\cite{Jensen:2004}. This hydrophobic matching principle has been very useful in explaining a number of effects (e.g., the transport properties of membrane channels), but is only a partial description. Even in the continuum limit, solving the elasticity equation requires a second piece of information, namely the slope imposed to the membrane thickness. The importance of the imposed slope is shown, for instance, by its influence on the lifetime of the channel formed by gramicidin, a widely studied antimicrobial peptide~\cite{Goulian:1998,Nielsen:1998}.

Although most of the experimental techniques (and of the theoretical approaches) consider the membrane inclusions as isolated objects, the study of dense systems by scattering techniques~\cite{He:1994,Harroun:1999} can yield significant information at the nanometer scale, impossible to obtain by other means. For instance, using small-angle x-ray scattering (SAXS) one can measure the structure factor $S(q)$ of the system as a function of the scattering vector $q$ and determine from it the interaction potential $V(d)$ between two inclusions as a function of the distance $d$ between them~\cite{Constantin:2007,Constantin:2008,Constantin:2009,Constantin:2010,Pansu:2011}. 

We consider the interaction potential between gramicidin channels inserted in membranes with two compositions: either lipids with a phosphocholine head group, which are major components of biological membranes, or single-chain nonionic surfactants with elastic properties very different from those of the lipids. For our study, we use 1,2-dilauroyl-\textit{sn}-glycero-3-phosphocholine (DLPC) and pentaethylene glycol monododecyl ether (C$_{12}$E$_5$), respectively. We calculate $V(d)$ by a continuum elasticity model~\cite{Bitbol:2012} in terms of the material parameters for the channel and the membranes (relevant lengths and elastic moduli), and of the coupling parameters. We show that the channels impose to the membrane a pronounced downward slope. We confirm this conclusion by applying our model to other data in the literature.

\begin{figure}
\includegraphics[width=0.9\columnwidth]{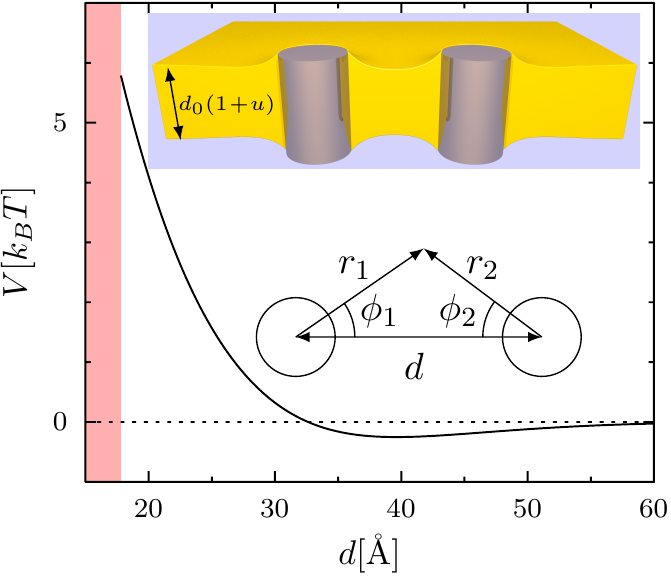}
\caption{Interaction energy~$V$ between two gramicidin channels in a DLPC bilayer as a function of their center-to-center distance~$d$. The red shading visualizes the contact between the inclusions. The parameters of the model correspond to our best fit of the experimental data. Upper inset: corresponding shape of the membrane for~$d=2 d_0 = \SI{43}{\angstrom}$. The channels are represented as cylinders. The relative deformation~$u$ is magnified by a factor of~$3$. Lower inset: coordinates for the multipolar expansion.\label{fig:edlpc}}
\end{figure}

We consider a tensionless bilayer membrane undergoing symmetric thickness variations around a flat midsurface (see Fig.~\ref{fig:edlpc}). To second order in the membrane excess thickness $u$ and its gradients, the most general expression for the deformation free-energy density is \cite{SM,Bitbol:2012}
\begin{align}
f &= \frac{1}{2} u^2 + \frac{k_1}{2} (\nabla u)^2 + \frac{k_2}{2} (\nabla^2 u)^2
\nonumber\\
&+ a_1 \nabla^2 u + a_2 \nabla \cdot (u \nabla u) + \bar{k} \det
(\nabla\nabla u).
\label{eq:f}
\end{align}
This free energy density is normalized by the bilayer compressibility modulus $K_a$~\cite{Gil:1998} and all lengths (including $u$) are normalized by the equilibrium thickness $d_0$. The dimensionless elastic constants in Eq.~(\ref{eq:f}) are related to the usual elastic constants as follows~\cite{Bitbol:2012}: $k_2=\kappa_0/(4K_a d_0^2)$ is proportional to the monolayer bending rigidity $\kappa_0/2$; $a_1=\kappa_0c_0/(2K_a d_0)$ to the monolayer spontaneous curvature $c_0$; $a_2=\kappa_0(c_0-c_0'\Sigma_0)/(2K_a d_0)$, where $c_0'$ is the derivative of the spontaneous curvature $c_0$ with respect to the molecular area $\Sigma_0$; $\bar k=\bar\kappa/(4K_a d_0^2)$, where $\bar\kappa$ is twice the Gaussian modulus of the monolayer; $k_1=K'_a/K_a$ is proportional to the tension-like parameter $K'_a$, which is unknown and in principle non-negligible as it reflects the energy cost associated with the gradients of the area per molecule, not accounted for by the other terms~\cite{Bitbol:2012}.

The cross-sectional area of the gramicidin channel is $A_0=\SI{250}{\angstrom^2}$ \cite{Harroun:1999}, corresponding to a hard core radius $R_0=\sqrt{A_0/\pi}=\SI{8.9}{\angstrom}$. Its thickness is $h_G=\SI{23}{\angstrom}$ \cite{Elliott:1983}. For DLPC (C12:0) membranes we estimate the following values of the elastic parameters: $K_a=\SI{0.235}{\newton\per\meter}$ and $\kappa_0=\SI{5.6e-20}{\joule}$ (from the values for C13:0 and C14:0 in Table~1 of Ref.~\citenum{Marsh:2006}), $\bar\kappa=-0.8 \kappa_0=\SI{-4.48e-20}{\joule}$ (see \S4 of Ref.~\citenum{Marsh:2006}), $d_0=\SI{20.8}{\angstrom}$\cite{Harroun:1999}, $c_0=\SI{-0.005}{\per\angstrom}$\cite{Templer:1994}, $c'_0\simeq \SI{0}{\angstrom}$. The area per lipid molecule is $A_l = \SI{63.2}{\angstrom^2}$ \cite{Kucerka:2005}. The excess hydrophobic thickness $U_0=h_G - d_0 = \SI{2.2}{\angstrom}$. For C$_{12}$E$_5$ membranes we use~\cite{Sottmann:1997,Kurtisovski:2007}: $K_a=\SI{0.25}{\newton\per\meter}$, $\kappa_0=\SI{8e-21}{\joule}$, $\bar\kappa=\SI{-3.04e-21}{\joule}$, $c_0=\SI{0.0266}{\per\angstrom}$, $c'_0\simeq \SI{0}{\angstrom}$, $A_l = \SI{42.9}{\angstrom^2}$. We estimate $d_0=\SI{16.9}{\angstrom}$ from the volume of a dodecyl chain compared to the molecular volume of C$_{12}$E$_5$, yielding $U_0=h_G - d_0=\SI{6.1}{\angstrom}$. All measurements were performed at $21 ^{\circ}\, \text{C}$.
 
Because of hydrophobic matching, each inclusion imposes a fixed excess thickness $u(r_0,\phi)=u_0$ along its boundary $r=r_0$, where $(r,\phi)$ are polar coordinates centered on the inclusion. Since the elastic energy includes second-order derivatives, the equilibrium thickness profile also depends on the radial derivative of $u$ along the boundary. We therefore assume that the inclusions set a preferred angle via a quadratic boundary potential (per unit length), which can be written in two equivalent forms: 

\begin{equation}
g = \frac{w}{2}(\partial_r u|_{r_0} -s)^2 = -\tau\,\partial_r u|_{r_0} +\frac{w}{2}(\partial_r u|_{r_0})^2+\text{cst.}
\label{eq:ws}
\end{equation}

\noindent where $w$ is an anchoring strength, $s$ the tangent of the preferred angle, and $\tau\equiv ws$ is the torque exerted by the inclusion when the boundary angle vanishes. According to the magnitude of $w$, we can distinguish three cases: (i) weak anchoring, where one can set $w=0$ without changing substantially the membrane profile $u$, (ii) strong anchoring, which amounts to letting $w \rightarrow \infty$ in \eqref{eq:ws}, and (iii) intermediate strength, where $\partial_r u|_{r_0}$ is generally different from $s$, but the effect of the anchoring on the profile is considerable. Only cases (i) and (ii) were used in the literature~\cite{Huang:1986,Helfrich:1990,Goulian:1998}, with various values of $s$.

Calculating the total elastic free energy via multipole expansion \cite{SM} we obtain the interaction potential $V(d)$. By Monte Carlo simulation, we follow the positions $\bm{r_i}$ of $N\simeq1000$ hard-core particles, interacting with the pairwise potential $V(d)$ in a confining circular box of radius~$r_\text{box}=40$. We compute the structure factor $S(q)=N^{-1}\langle\sum_{i,j=1}^N\!J_0(q|\bm{r}_i-\bm{r}_j|)\rangle$, where $J_0$ is the Bessel function, averaging over all the realizations and all the directions of the wave vector $\bm q$, for liquidlike ordering.

Among the unknown material constants, $a_1$, $a_2$ and $\bar{\kappa}$ have negligible effect on the interaction potential, as checked by extensive simulations. We adjust the remaining parameters $k_1$, $w$ and~$s$ by a global fitting procedure to the set of seven experimental spectra~$S(q)$ measured in Ref.~\citenum{Constantin:2009} for different gramicidin concentrations in the same experimental conditions.

\begin{figure}[htbp]
\includegraphics[width=0.9\columnwidth]{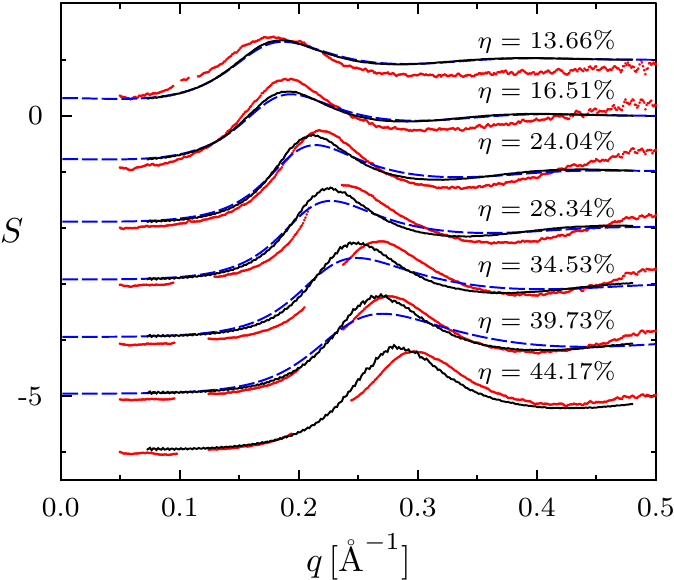}
\caption{Structure factors~$S$ as a function of the scattering vector~$q$ for gramicidin channels in DLPC bilayers at different surface fractions~$\eta$ of inclusions. Red dots: experimental data from Ref.~\citenum{Constantin:2009}. Black curves: Monte Carlo simulations for the parameters of the model corresponding to our best fit of the experimental data (see text). Blue dashed lines: HNC approximations. Curves are shifted downwards by unit increments.
\label{fig:sdlpc}}
\end{figure}

For DLPC, $V(d)$ is shown in Fig.~\ref{fig:edlpc}. It exhibits a relatively short-ranged repulsion of a few $k_B T$ close to contact, followed by a shallow attractive well. In the same Fig.~\ref{fig:edlpc}, we show the calculated shape of the membrane for a distance $d=2d_0=\SI{43}{\angstrom}$.
%Notice that in-between the two inclusions the membrane shrinks below its equilibrium thickness; in the outer region, the membrane relaxes towards its equilibrium thickness with damped oscillations (in the figure, only the first oscillation is visible).
%For the shown distance, the dip in-between inclusions is deeper than the first outer oscillation; for inclusions close to contact, the dip persists but becomes shallower. 

In the outer region the contact angle $\arctan(\partial_r u|_{r_0}) \approx \partial_r u|_{r_0}$ is~$\simeq -37^\circ$ and is fixed by the competition between the torque exerted by the inclusion and the elastic deformation induced on the membrane.

Figure~\ref{fig:sdlpc} displays $S(q)$ as a function of the scattering vector~$q$: experimental data of Ref.~\citenum{Constantin:2009} (red dots) and corresponding Monte Carlo results, computed for~$6\times 10^6$ Monte Carlo steps after thermal equilibration (black curves). A best fit is found for $k_1=\num{5.4+-3e-2}$, $\tau=\num{-4.7+-0.4e-2}$, $w<\num{1e-3}$: see Fig.~\ref{fig:sdlpc}. Assuming a constant uncertainty $\sigma_S=0.1$ for the experimental points~\cite{Constantin:2009}, the goodness-of-fit function $\chi^2$ is of the order of $2.1$ per data point.

We also tested the approximate hypernetted-chain (HNC) solution, shown by dashed blue lines in Fig.~\ref{fig:sdlpc}. The discrepancies with respect to the Monte Carlo results as the surface fraction $\eta$ increases can be explained by the diffuse character of the repulsion, together with the presence of the attractive well~\cite{Hus:2013}.
%Also, at the highest surface fraction, the Lado algorithm did not converge. 

\begin{figure}[t]
\includegraphics[width=0.9\columnwidth]{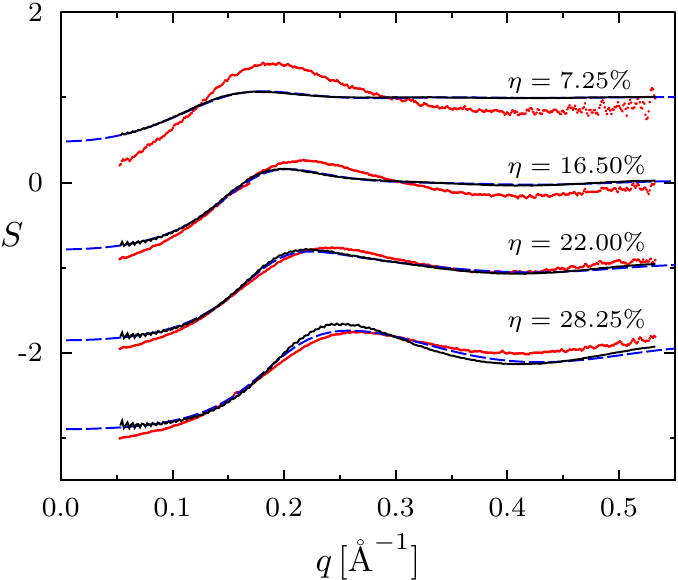}	
\caption{Experimental structure factors and fits for for gramicidin channels in C$_{12}$E$_5$ bilayers. Same notation as in Fig.~\ref{fig:sdlpc}.  \label{fig:c12e5}}
\end{figure}

\begin{figure}[t]
\includegraphics[width=0.9\columnwidth]{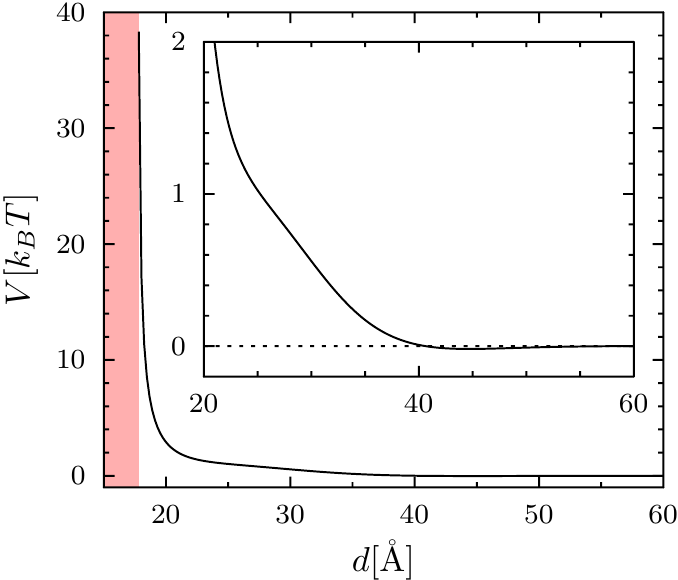}	
\caption{Interaction energy $V$ between two gramicidin channels in
C$_{12}$E$_5$ bilayers for the model parameters corresponding to our best fit of the experimental data. Shaded area: contact between the inclusions. Inset: zoom for $d>20 \text{\AA}$.
\label{fig:ec12e5}}
\end{figure}

For C$_{12}$E$_5$ we used the same global fitting procedure as for DLPC. The experimental spectra~$S(q)$ measured in Ref.~\citenum{Constantin:2009} yield the best fit shown in Fig.~\ref{fig:c12e5} for $k_1=\num{1.04+-0.8e-2}$, $w=\num{0.85+-0.15}$, $s=\num{-1.23+-0.1}$. Assuming again a constant uncertainty $\sigma_S=0.1$ for the experimental points~\cite{Constantin:2009}, the goodness-of-fit function $\chi^2$ is of the order of $1.3$ per data point. The interaction potential $V(d)$ is shown in Fig.~\ref{fig:ec12e5}. The equilibrium slope for an isolated inclusion is close to the preferred value~$s$, corresponding to an angle~$\simeq-50^\circ$. Although this value is outside the validity range of our linearized model, it clearly indicates a large negative angle.

The hydrophobic matching principle is very general: due to the high cost of exposing hydrophobic residues to water and to the large difference in compression moduli between proteins and membranes~\cite{Jensen:2004}, the contact hydrophobic thickness of any bilayer equals that of the embedded protein. With the same generality, we assume that the boundary condition for the slope, being set locally at the contact between the protein surface and the hydrophobic/hydrophilic interface of the bilayer, only depends on the nature of the inclusion and on the chemical family of the lipids (defined by the nature of the polar head).

Gramicidin must then impose the same torque $\mathcal{T}=K_ad_0\tau$ on all lipids with a phosphocholine (PC) head, irrespective of the length of the alkyl chain~\footnote{Which will, however, affect other parameters of the bilayer, such as the elastic moduli.}, yielding negative equilibrium boundary angles of similar magnitudes. One counterintuitive consequence is that, for PC membranes with a hydrophobic length larger than that of the gramicidin, the deformation profile should decrease steeply and then increase back to its equilibrium value.

The hydrophobic length of the gramicidin channel is larger than that of DLPC and C$_{12}$E$_5$ bilayers and it is not surprising that, in both cases, it imposes a preferential steep negative slope. To validate the generalization above, one would also need to demonstrate a negative contact slope in thicker bilayers. No SAXS results are available for phospholipids with longer chains, but other types of experimental and numerical results are available in the literature. In the following, we show that they support our hypothesis.

Several authors have simulated gramicidin channels inserted within thicker lipid bilayers~\cite{Woolf:1996,Chiu:1999,Allen:2004,Yoo:2013}, and their results could provide evidence as to the membrane profile.  Three publications \cite{Chiu:1999,Yoo:2013,Beaven:2017} do present such data (see the Supplemental Material \cite{SM} for more details) and support our conclusion of a steep decrease in the thickness of phosphocholine membranes at the boundary with the gramicidin channel even for thicker bilayers, resulting in a nonmonotonic profile.

\begin{figure}[t]
\includegraphics[width=0.9\columnwidth]{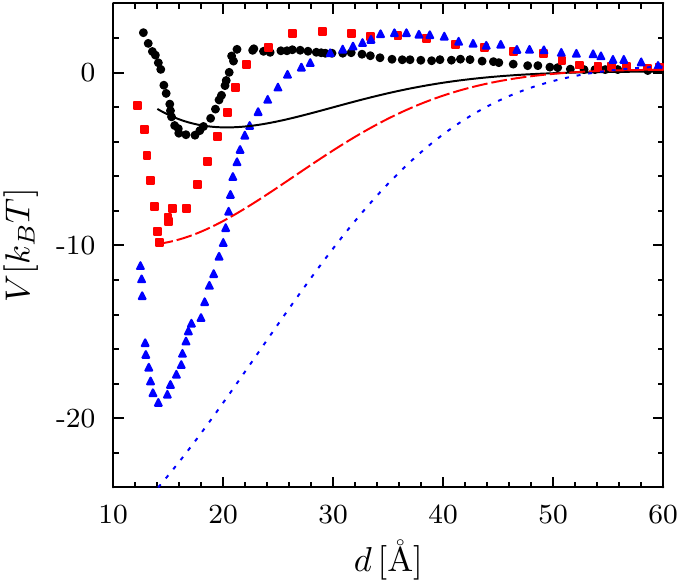}	
\caption{PMF between two gramicidin channels, simulated by Yoo and Cui (symbols; redrawn from Fig.~2A of Ref.~\citenum{Yoo:2013b}) and calculated using the present model (lines) in DMPC (black solid dots and solid line), DPPC (red squares and dashed line) and DSPC (blue triangles and dotted line) bilayers, with no adjustable parameters.
\label{fig:Yoo}}
\end{figure}

Using coarse-grained molecular dynamics, Yoo and Cui simulated two channels embedded into bilayers with different compositions (DMPC, DPPC or DSPC)~\cite{Yoo:2013b}, computed the potential of mean force (PMF) and found in each case similar behavior, consisting of a steep short-range attraction and a long-range repulsion. In Fig.~\ref{fig:Yoo} we present their results and compare them to the predictions of our elastic model. No fitting is involved: we keep $K'_a$ and~$\mathcal{T}$ fixed at the values obtained above for DLPC bilayers, in agreement with our assumption, and we use literature data \cite{SM} for the other material constants of the three lipid systems (notably, the parameters $K_a$ and $d_0$ that appear in the normalized constants $k_1$ and~$w$). The hard core radius is taken as $R_0 = \SI{7}{\angstrom}$, for coherence with Ref.~\citenum{Yoo:2013b}. Our model yields both the attractive and the repulsive part, although shifted further away from contact, possibly due to the inherent approximation of the coarse-grained simulation model or to higher-order gradient terms. Moreover, the contact value of the predicted potential is in very good agreement with the simulations.

\begin{figure}[t]
\includegraphics[width=0.9\columnwidth]{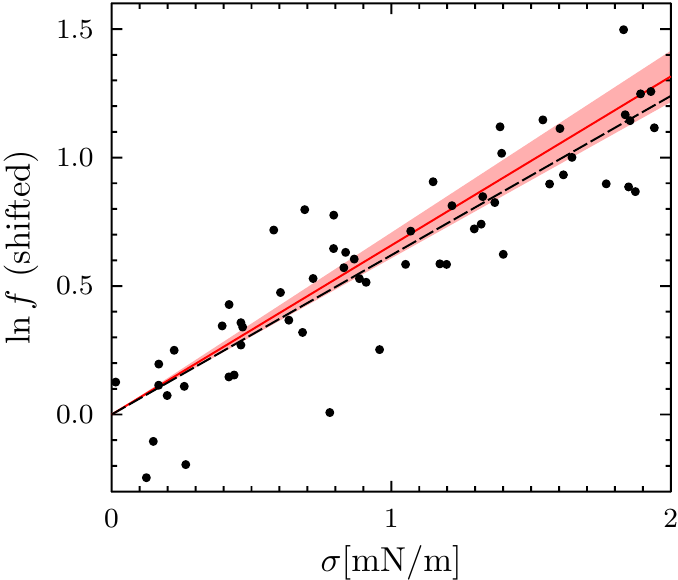}	
\caption{Logarithm of the formation rate $f$ of gramicidin channels in
DOPC bilayers as a function of the membrane tension $\sigma$. Solid
dots: experimental data retrieved from Fig.~6b of
Ref.~\citenum{Goulian:1998} (after  subtraction of a constant baseline,
see Ref~\citenum{Bitbol:2012} for the detailed analysis). Solid red
line and red shaded area: our prediction using the elastic model and the
parameter values discussed in the text (with no adjustable parameters).
Black dashed line: best linear fit.
\label{fig:sigma}}
\end{figure}

Gramicidin is widely used due to its conductivity properties: the channel switches between the open (ion-conducting) and closed states as the monomers dimerize and dissociate, respectively. The transition between states can be followed by conductivity measurements~\cite{Hwang:2003} as a function of an externally controlled parameter, as shown for gramicidin channels in DOPC under a variable applied tension $\sigma$~\cite{Goulian:1998}. In particular, the formation rate $f$ of the channels can be described, for moderate tension $\sigma < \SI{2}{mN/m}$, as
\begin{equation}
\label{eq:fsigma}
\ln f = C_0 + C_1 \sigma,
\end{equation}
\noindent where $C_0$ changes with the details of the experimental situation, while $C_1$ only depends on the intrinsic elastic parameters \cite{SM}. Based on our assumption that all PC lipids share the same anchoring properties, we use the parameter values obtained by fitting the $S(q)$ data for channels in DLPC bilayers, which yield for DOPC (after proper normalization with $K_a$ and $d_0$), $k_1 = \num{4.8e-2}$ and $\tau=\num{3.6e-2}$. Supplemented by literature data~\cite{Rawicz:2000,Goulian:1998,Szule:2002} for the other material constants of DOPC, these parameters yield a slope $C_1 =\SI{658+-50}{m/N}$, in excellent agreement with the best fit to the experimental data $C^{\text{exp}}_1  = \SI{620}{m/N}$ (see Fig.~\ref{fig:sigma}). This correspondence is striking since parameters extracted from the SAXS data are used to predict conductivity results, showing that both kinds of measurements are well captured by the continuum elastic model.

A nonmonotonic interface profile close to the inclusion was invoked more than thirty years ago by Huang~\cite{Huang:1986}, in order to explain the results of Elliott \textit{et al.}~\cite{Elliott:1983} concerning the lifetime of gramicidin channels in monoacylglycerol bilayers. Although they concern membranes of different composition, our results indirectly confirm Huang's insight and emphasize the role of higher-order terms in membrane elasticity.

To summarize, we present a complete elastic model for the membrane-induced interaction between inclusions, comprising a boundary energy associated to the slope of the membrane thickness at its contact
with the inclusion. By fitting experimental results for the interaction
of gramicidin channels in two types of bilayers we obtain the first
quantitative measurement of the preferred slope and the associated
torque or stiffness constants and show that these parameters are essential for a realistic description of the inclusion-membrane interaction in terms of continuum elasticity.

We confirm our results and extend the analysis to phospholipids with
longer chains by applying it to the potential of mean force between
channels obtained by simulations~\cite{Yoo:2013b} and to data on the
conductivity of the channel under tension~\cite{Goulian:1998}.
Surprisingly, even when the hydrophobic thickness of a phospholipid
bilayer is larger than that of the gramicidin channel, the membrane
thickness first decreases with the distance from the boundary before
increasing to its equilibrium value far away from the inclusion.

The strongly negative value of the thickness slope $s$ might result
from the combination between the specific interactions at the contact of
the protein with the hydrophobic/hydrophilic interface of the bilayer
and the conical shape of the gramicidin monomer~\cite{Brasseur:1986},
which gives the channel an hourglass shape rather than a cylindrical
one. If, due to the local interactions, the molecules neighboring the
channel tend to be parallel to the sides of the monomers, their axis
(and hence the normal to the monolayers) is tilted away from the
vertical, resulting in a negative $s$ for a wide variety of membrane
components. This could explain the very similar value of $s$ obtained in
bilayers formed by C$_{12}$E$_5$ (see above), whose chemical nature is
quite different from that of PC lipids.

\begin{acknowledgments}
We acknowledge financial support from the French Agence
Nationale de la Recherche (Contract No.\ ANR-12-BS04-0023-MEMINT) and
useful discussions with Anne-Florence Bitbol. The SAXS experiments were performed on beam line D2AM at the European Synchrotron Radiation Facility (ESRF), Grenoble, France. We are grateful to Cyrille Rochas for providing assistance in using beam line D2AM.
\end{acknowledgments}

%\bibliography{Bories}


\end{document}